\documentclass[twocolumn,showpacs,preprintnumbers,amsmath,amssymb]{revtex4}

\usepackage{graphicx}
\usepackage{dcolumn}
\usepackage{bm}

\begin{document}

\preprint{Submission to Phys. Rev. B}

\title{
Magnetism of the antiferromagnetic spin-$\frac{1}{2}$ tetramer compound 
CuInVO$_5$ 
}

\author{Masashi Hase$^1$}
 \email{HASE.Masashi@nims.go.jp}
\author{Masashige Matsumoto$^2$}
\author{Akira Matsuo$^3$}
\author{Koichi Kindo$^3$}

\affiliation{%
${}^{1}$National Institute for Materials Science (NIMS), 
Tsukuba, Ibaraki 305-0047, Japan \\
${}^{2}$Department of Physics, Shizuoka University, 
Shizuoka 422-8529, Japan \\
${}^{3}$The Institute for Solid State Physics (ISSP), 
the University of Tokyo, 
Kashiwa, Chiba 277-8581, Japan
}%

\date{\today}

\begin{abstract}

We measured the temperature dependence of 
the magnetic susceptibility and specific heat 
and the magnetic-field dependence of 
the magnetization of CuInVO$_5$. 
An antiferromagnetically ordered state
appears below $T_{\rm N} = 2.7$ K. 
We observed a $\frac{1}{2}$ quantum magnetization plateau 
above 30 T at 1.3 K. 
We show that 
the spin system consists of 
antiferromagnetic spin-$\frac{1}{2}$ tetramers 
with $J_1 = 240 \pm 20$ and $J_2 = -142 \pm 10$ K 
for the intratetramer interactions. 

\end{abstract}

\pacs{75.10.Jm, 75.50.Ee, 75.40.Cx, 75.47.Lx}

\maketitle

\section{INTRODUCTION}

The interacting antiferromagnetic (AF) spin-dimer compounds 
TlCuCl$_3$ \cite{Tanaka03,Oosawa04,Oosawa99,Nikuni00,Tanaka01} 
and 
KCuCl$_3$ \cite{Goto06,Oosawa02} 
show a pressure-induced or magnetic-field-induced 
magnetic quantum phase transition. 
Experimental observations \cite{Kuroe08,Ruegg08,Merchant14,Kuroe12}
and 
the theoretical background \cite{Matsumoto04,Matsumoto08a}
of massive longitudinal-mode magnetic excitations 
in the ordered state were reported for these compounds.  
The longitudinal mode and 
massless transverse modes (Nambu-Goldstone modes) \cite{Goldstone62} 
are related to fluctuations in 
the amplitude and phase of the order parameter, respectively. 
The longitudinal mode 
is the analog of the Higgs particle \cite{Higgs64,Sachdev11,Podolsky11}. 

According to the results of theoretical investigations, 
the longitudinal mode can exist in 
interacting AF spin cluster systems that are realized in 
Cu$_2$Fe$_2$Ge$_4$O$_{13}$ and Cu$_2$CdB$_2$O$_6$ \cite{Matsumoto10}. 
The spin systems in 
Cu$_2$Fe$_2$Ge$_4$O$_{13}$ \cite{Matsumoto10}
and 
Cu$_2$CdB$_2$O$_6$ \cite{Hase05,Hase09,Hase15}
can be regarded as 
interacting AF spin tetramers  
(Fe-Cu-Cu-Fe and Cu-Cu-Cu-Cu tetramers, respectively). 
The shrinkage of ordered magnetic moments by quantum fluctuation is important 
for the appearance of the longitudinal mode.   
The ground state (GS) can be a spin-singlet state 
in isolated AF spin clusters. 
Therefore, some interacting spin cluster systems 
are advantageous for the longitudinal mode. 
An antiferromagnetically ordered state appears 
in Cu$_2$Fe$_2$Ge$_4$O$_{13}$ \cite{Masuda04} and 
Cu$_2$CdB$_2$O$_6$ \cite{Hase05}
in zero magnetic field under atmospheric pressure. 
The magnetic excitations in Cu$_2$Fe$_2$Ge$_4$O$_{13}$ 
have been investigated by inelastic neutron scattering (INS) experiments 
on single crystals \cite{Masuda04,Masuda05,Masuda07,Masuda09}. 
The longitudinal mode was not confirmed 
because of the small INS intensities 
due to the large excitation energies ($> 15$ \ meV) and 
because of the overlap of the transverse modes. 
The magnetic excitations in Cu$_2$$^{114}$Cd$^{11}$B$_2$O$_6$ 
were studied by INS experiments on its powder \cite{Hase15}. 
Although the results suggest the existence of the longitudinal mode, 
there was no conclusive evidence 
because powder was used. 
A single crystal suitable for the measurements of physical properties 
has not been reported. 

We require further spin cluster compounds that have 
an antiferromagnetically ordered state and 
low-energy longitudinal-mode magnetic excitations. 
We focus on spin-$\frac{1}{2}$ tetramers 
because of the following magnetism. 
The Hamiltonian of a spin tetramer is expressed as 
\begin{equation}
{\cal H} = 
J_1 S_{2} \cdot S_{3} + J_2 (S_{1} \cdot S_{2}+ S_{3} \cdot S_{4}).
\end{equation}
When $J_1 >0$ or $J_2 >0$, 
the GS is the spin-singlet state. 
Therefore, the shrinkage of ordered moments can be expected 
in an ordered state generated by 
the introduction of intertetramer interactions. 
The ordered state is possible under the condition that 
the value of $\Delta$ is comparable to or less than 
that of an effective intercluster interaction \cite{Matsumoto10}.   
Here $\Delta$ is the energy difference (spin gap)
between the singlet GS and first-excited triplet states.  
The effective intercluster interaction is given by the sum of 
the products of the absolute value of each intercluster interaction 
($|J_{{\rm int}, i}|$) and 
the corresponding number of interactions per spin ($z_i$) 
as $J_{\rm eff} = \sum_i z_i |J_{{\rm int}, i}|$. 
The effective intercluster interaction is usually much smaller 
than the dominant intracluster interactions.  
Therefore, $\Delta$ should be much smaller 
than the dominant intracluster interactions 
for the appearance of the ordered state.  
Figure 1 shows 
the eigenenergies of the excited states measured from the GS 
in an isolated spin-$\frac{1}{2}$ tetramer \cite{Hase97}. 
As shown in Fig. 1(a) for $J_1 >0$, 
$\Delta / J_1$ can be sufficiently small 
when $J_2$ has negative or small positive values. 
Even under a small $J_{\rm eff}$,
an ordered state is expected in a spin-tetramer compound for $J_1 >0$ and $J_2 <0$.
The small $\Delta / J_1$ is in contrast to 
$\Delta /J = 1 $ 
in the AF spin-$\frac{1}{2}$ dimer given by $J S_1 \cdot S_2$. 
As shown in Fig. 1(a), 
the GS and first-excited states are well separated from 
the other excited states (ESs). 
This means that 
the low-energy physics can be described by 
an effective spin-dimer (singlet-triplet) system 
\cite{Matsumoto10}.

\begin{figure}
\begin{center}
\includegraphics[width=7cm]{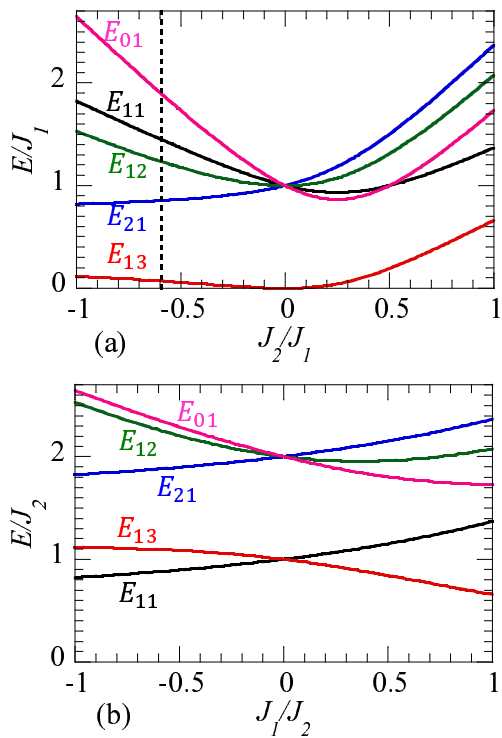}
\caption{
(Color online)
Eigenenergies of excited states measured 
from the ground state in 
an isolated spin-$\frac{1}{2}$ tetramer expressed by Eq. (1).  
There are 
two $S^{\rm T}=0$ states ($|01>$ and $|02>$), 
three $S^{\rm T}=1$ states ($|11>$, $|12>$, and $|13>$), and 
one $S^{\rm T}=2$ state ($|21>$). 
$S^{\rm T}$ is 
the value of the sum of the spin operators in the tetramer. 
The eigenstates $|ij \rangle$ of  the isolated tetramer
are explicitly given in \cite{Hase97}. 
In the isolated tetramer, 
the  ground state is the spin-singlet $|02 \rangle$ state.
(a) $J_1 >0$. 
The vertical dashed line indicates the $J_2 / J_1$ value of 
CuInVO$_5$ evaluated in the present work.
(b) $J_2 >0$. 
}
\end{center}
\end{figure}

We can expect spin-$\frac{1}{2}$ tetramers in CuInVO$_5$ 
from its crystal structure \cite{Moser99}.
The Cu$^{2+}$ ions ($3d^9$) have localized spin-$\frac{1}{2}$. 
The positions of the Cu ions and the O ions connected to 
the Cu ions are shown schematically 
in Fig. 2(a). 
Two crystallographic Cu sites (Cu1 and Cu2) exist. 
Red and blue bars indicate 
the shortest and second-shortest Cu-Cu bonds, respectively.
The Cu-Cu distances at room temperature are 
3.117 and 3.173 \AA, respectively. 
The shortest bond has 
two identical Cu1-O-Cu1 paths whose angle is $89.75^{\circ}$. 
The second-shortest bond has two different Cu1-O-Cu2 paths with  
angles of 107.61 and 88.19$^{\circ}$. 
The Cu-Cu distances in the other bonds are 4.705 \AA \ or greater.  
If dominant exchange interactions exist in 
the shortest and second-shortest Cu-Cu bonds, 
spin tetramers given by Eq. (1) are formed. 
Figure 2(b) shows the arrangement of the spin tetramers. 
Two types of tetramers (I and II) exist, 
although they are equivalent to each other 
as a spin system. 
In this paper, 
we report the magnetism of CuInVO$_5$. 
An AF long-range order  
appears below $T_{\rm N} = 2.7$~K.
We show that 
the spin system consists of spin tetramers 
with $J_1 >0$ and $J_2 <0$. 

\begin{figure}
\begin{center}
\includegraphics[width=7cm]{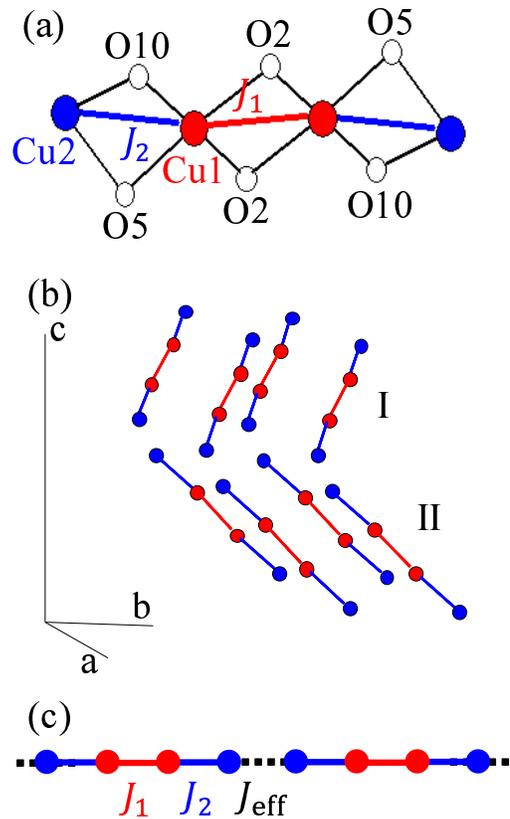}
\caption{
(Color online)
(a)
Schematic drawing of positions of 
Cu$^{2+}$ ions having spin-$\frac{1}{2}$ and 
O$^{2-}$ ions connected to Cu$^{2+}$ ions  
in CuInVO$_5$. 
Red, blue, and white circles indicate Cu1, Cu2, and O sites, respectively. 
Red and blue bars represent 
the shortest and second-shortest Cu-Cu bonds, respectively. 
Thin black bars represent Cu-O bonds. 
We define $J_1$ and $J_2$
as the exchange interaction parameters for 
the shortest and second-shortest Cu-Cu bonds, respectively. 
The $J_1$ and $J_2$ interactions form 
a spin-$\frac{1}{2}$ tetramer. 
(b)
Schematic drawing of spin tetramers 
in CuInVO$_5$. 
Two types of tetramers (I and II) exist, 
although they are equivalent to each other 
as a spin system. 
The space group of CuInVO$_5$ is $P2_1/c$ (No. 14). 
The lattice constants at room temperature are 
$a=8.793(2)$, $b=6.1542(6)$, $c=15.262(2)$ \AA, 
$\beta=106.69(2)^{\circ}$, 
and $Z=8$ (8 formulas per unit cell) \cite{Moser99}. 
(c)
Interacting spin tetramer model used to calculate magnetization 
using a mean-field theory based on the tetramer unit 
(tetramer mean-field theory). 
}
\end{center}
\end{figure}

\section{Experimental and Calculation Methods}

Crystalline CuInVO$_5$ powder was synthesized 
by a solid-state reaction  
at 1,023 K in air for 100 h with intermediate grindings. 
We confirmed the formation of CuInVO$_5$ 
using an x-ray diffractometer (RINT-TTR III; Rigaku).
We measured the specific heat using  
a Physical Property Measurement System (PPMS; Quantum Design). 
We measured the magnetization in magnetic fields of up to 5 T 
using a superconducting quantum interference device 
(SQUID) magnetometer 
Magnetic Property Measurement System (MPMS; Quantum Design). 
High-field magnetization measurements were conducted 
using an induction method with a multilayer pulsed field magnet
installed at the Institute for Solid State Physics (ISSP), 
the University of Tokyo. 

We obtained the eigenenergies and eigenstates of 
isolated spin-$\frac{1}{2}$ tetramers   
using an exact diagonalization method \cite{Hase97}. 
We calculated 
the temperature $T$ dependence of the magnetic susceptibility and 
the magnetic-field $H$ dependence of the magnetization $M(H)$  
using the eigenenergies and eigenstates. 
We calculated $M(H)$ for the model shown in Fig. 2(c) 
using a mean-field theory based on the tetramer unit 
(tetramer mean-field theory).
Finite magnetic moments were initially assumed 
on the Cu sites in the tetramer.
The mean-field Hamiltonian was then expressed by a $16 \times 16$ matrix form
under consideration of the external magnetic field and 
the molecular field from the nearest-neighbor sites. 
The eigenstates of the mean-field Hamiltonian were used 
to calculate the expectation value of the ordered moments on the Cu sites.
We continued this procedure until the values of the magnetic moments converged.
We finally obtained a self-consistently determined solution for $M(H)$. 

\section{Results and discussion}


The red circles in Figs. 3 and 4 show the $T$ dependence of 
the specific heat $C(T)$ of CuInVO$_5$
in zero magnetic field and 
the magnetic susceptibility $\chi (T)$ 
in a magnetic field of $H = 0.01$~T, respectively.  
We can observe a peak in $C(T)$ at 2.7 K and 
a clear decrease in $\chi (T)$ below this temperature, 
indicating the occurrence of an AF long-range order. 
A broad maximum can be seen 
around 8 K in $C(T)$ and around 11 K in $\chi (T)$, 
indicating that the origin of the broad maximum in $C(T)$ 
is magnetic \cite{Comment1}. 
As $T$ is increased, $\chi (T)$ 
decreases rapidly up to $T = 40$ K then 
decreases slowly at higher temperatures. 
Other phase transitions were not observed 
in $C(T)$ and $\chi (T)$ below 300 K.  

\begin{figure}
\begin{center}
\includegraphics[width=7cm]{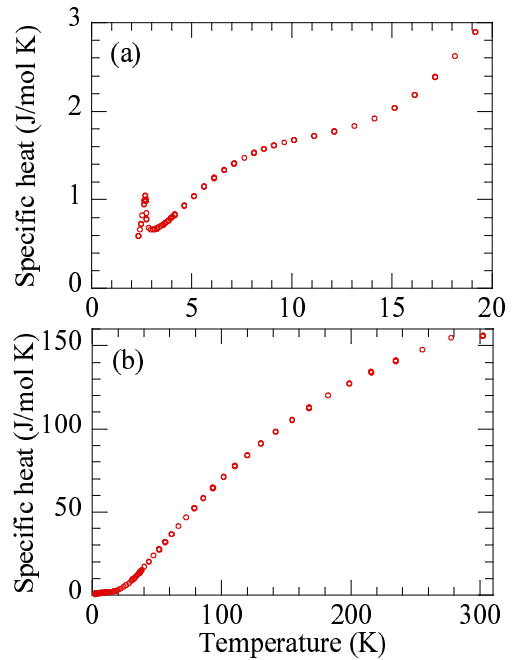}
\caption{
(Color online)
Temperature $T$ dependence of 
the specific heat $C(T)$ of CuInVO$_5$
in zero magnetic field.
}
\end{center}
\end{figure}

\begin{figure}
\begin{center}
\includegraphics[width=7cm]{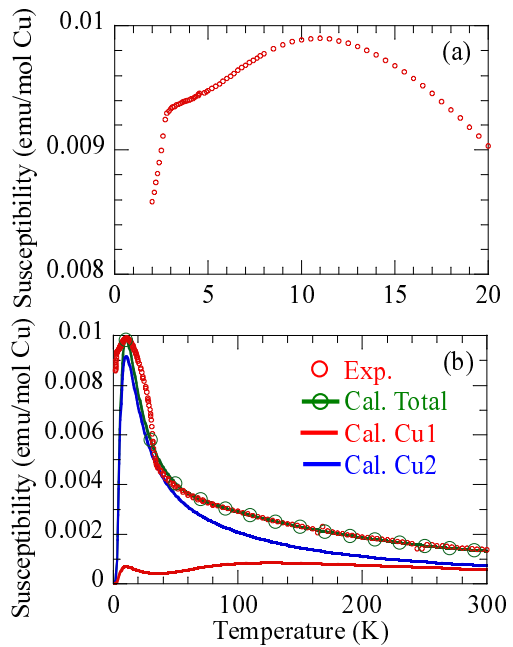}
\caption{
(Color online)
Temperature $T$ dependence of 
the magnetic susceptibility $\chi (T)$ of CuInVO$_5$ (circles)
in a magnetic field of $H = 0.01$ \ T. 
Green, red, and blue lines indicate 
$\chi (T)$ calculated for 
the total, Cu1, and Cu2 spins, respectively, 
in an isolated spin-$\frac{1}{2}$ tetramer.   
The $J_1$ and $J_2$ values are listed in Table I.
}
\end{center}
\end{figure}


The thick red lines in Figs. 5(a) and (b) show the $H$ dependence of 
the magnetization $M(H)$ of CuInVO$_5$ 
measured at 1.3 and 30 K, respectively. 
We can observe a $\frac{1}{2}$ quantum magnetization plateau 
above 30 T at 1.3 K. 
The $g$ value was evaluated to be $2.09 \pm 0.02$
from the magnetization of the plateau. 
The magnetization plateau is smeared at 30 K. 

\begin{figure}
\begin{center}
\includegraphics[width=7cm]{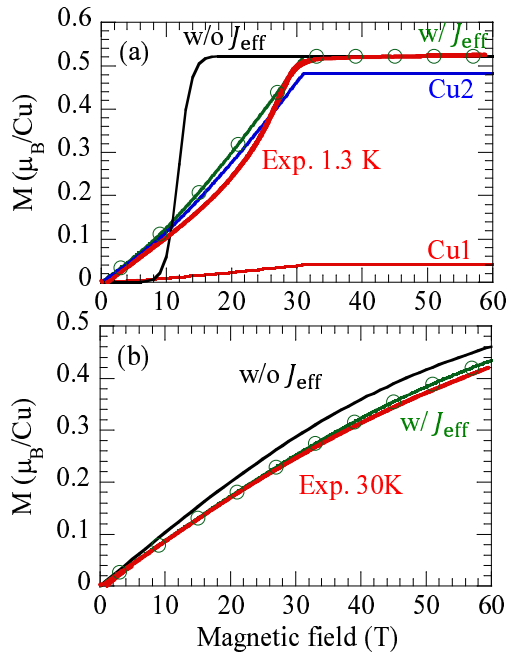}
\caption{
(Color online)
Magnetic-field dependence of the magnetization 
of CuInVO$_5$ (thick red lines). 
Green, red, and blue lines indicate 
the magnetization calculated for 
the total, Cu1, and Cu2 spins, respectively, 
in the interacting spin-$\frac{1}{2}$ tetramer model in Fig. 2(c). 
Black lines indicate 
the magnetization calculated for 
an isolated spin-$\frac{1}{2}$ tetramer. 
The values of the exchange interactions are listed in Table I.
(a)
Magnetization at 1.3 K. 
(b)
Magnetization at 30 K. 
}
\end{center}
\end{figure}


We compare $\chi (T)$ and $M(H)$ for CuInVO$_5$ with 
those calculated for isolated spin tetramers. 
The green line in Fig. 4(b) indicates 
$\chi (T)$ calculated for an isolated spin tetramer 
with $J_1 = 240$ and $J_2 = -142$ K. 
The $J_1$ and $J_2$ values are listed in Table I.  
The agreement between the experimental and calculated $\chi (T)$ 
is nearly perfect above 30 K, 
whereas a discrepancy  is seen below 30 K. 
The black lines in Fig. 5 indicate 
$M(H)$ calculated for an isolated spin tetramer 
with the same $J_1$ and $J_2$ values. 
The  calculated $M(H)$ is similar to the experimental $M(H)$ at 30 K, 
whereas 
the isolated spin tetramer model fails to reproduce 
the experimental $M(H)$ at 1.3 K. 

The agreement between the experimental and calculated results 
above 30 K indicates that 
the spin system in CuInVO$_5$ consists of 
spin tetramers with $J_1 = 240$ and $J_2 = -142$ K. 
To stabilize the ordered state, intertetramer interactions must exist in CuInVO$_5$.
Intertetramer interactions have a greater effect on the magnetization 
at lower $T$. 
Therefore, the discrepancy between the experimental results 
and those calculated for the isolated spin tetramer 
appears at low $T$. 
The magnetic structure of CuInVO$_5$ has not yet been reported. 
It is difficult to determine which intertetramer interactions are effective. 
Therefore, we assumed the simple model shown in Fig. 2(c) 
and calculated $M(H)$ using the tetramer mean-field theory. 
Since multiple intertetramer interactions are expected in CuInVO$_5$,
$J_{\rm eff}$ is the effective interaction between tetramers.
As described below, 
the magnetic moment on Cu1 sites is small 
in the spin tetramer with $J_1 = 240$ and $J_2 = -142$ K. 
Therefore, we assumed intertetramer interactions between Cu2 spins. 
The green lines in Fig. 5 indicate 
$M(H)$ calculated for the interacting spin tetramer 
with $J_1 = 240$, $J_2 = -142$, and $J_{\rm eff}=30$ K. 
The experimental and calculated magnetizations 
are in agreement with each other 
at both 1.3 and 30 K. 

\begin{table}
\caption{\label{table1}
Values of exchange interaction parameters and $g$ value. 
We used the central values for the calculations of 
the magnetic susceptibility in Fig. 4, 
the magnetization in Fig. 5, and 
the eigenenergies in Fig. 6. 
}
\begin{ruledtabular}
\begin{tabular}{cccc}
$J_1$ (K) & $J_2$ (K) & $J_{\rm eff}$ (K) & g\\
\hline
$240 \pm 20$ & $-142 \pm 10$ & $30 \pm 4$ & $2.09 \pm 0.02$\\
\end{tabular}
\end{ruledtabular}
\end{table}

Figure 6 shows
the eigenenergies of the excited states 
measured from the GS ($|02 \rangle$ state)  
in the isolated spin tetramer with $J_1 = 240$ and $J_2 = -142$ K. 
The first excited states 
are the spin-triplet $|13 \rangle$ states
located at $\Delta = 17$ K.  
The condition for the appearance of the ordered state 
($\Delta \leq J_{\rm eff}$) is satisfied. 
The second excited states 
are the spin-quintet $|21 \rangle$ states
located at 205 K.  
The large energy difference between the first and second ESs
generates the $\frac{1}{2}$ quantum magnetization plateau in Fig. 5(a).  

\begin{figure}
\begin{center}
\includegraphics[width=7cm]{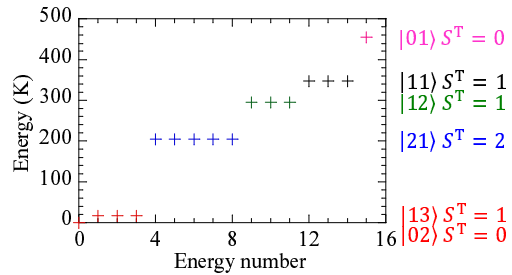}
\caption{
(Color online)
Eigenenergies of the excited states measured 
from the ground state ($|02>$ state) in 
the isolated spin-$\frac{1}{2}$ tetramer expressed by Eq. (1).  
The $J_1$ and $J_2$ values are listed in Table I.
}
\end{center}
\end{figure}


We roughly estimated the errors of the $J_1$, $J_2$, and $J_{\rm eff}$ values 
and listed them in Table I.
A discrepancy between the experimental and calculated $\chi (T)$ 
appears around 80 K when $J_1$ deviates from 240 K. 
The experimental and calculated $\chi (T)$ 
are not in agreement with each other when $J_1 = 220$ or 260 K. 
A discrepancy between the experimental and calculated $\chi (T)$ 
appears around 11 K when $J_2$ deviates from -142 K. 
The peak heights of the experimental and calculated $\chi (T)$ 
are not in agreement with each other when $J_2 = -132$ or -152 K. 
The magnetic field at which the 1/3 magnetization plateau appears 
increases with increasing $J_{\rm eff}$. 
We roughly estimated the error of $J_{\rm eff}$ to be $\pm 4$ K. 


The results calculated for spins on the Cu1 and Cu2 sites  
are shown in Figs. 4(b) and 5(a). 
We used the isolated and interacting spin tetramers in  
the calculations of $\chi (T)$ and $M(H)$, respectively. 
Cu2 spins show much larger magnetization than Cu1 spins at low $T$. 
The maximum $\chi (T)$ around 11 K and the rapid decrease up to $T = 40$ K  
mainly originate from the Cu2 spins. 
As $T$ is increased further, 
the susceptibility of Cu1 spins increases up to $T = 130$ K, 
whereas that of Cu2 spins decreases. 
Therefore, the total susceptibility shows weak $T$ dependence 
between 50 and 100 K. 
The most dominant interaction is the $J_1$ interaction. 
The spin state of Cu1 spins is similar to 
the singlet state in AF dimers \cite{Hase93a,Hase93b,Hase93c}. 
Therefore, the magnetization of Cu1 spins is small at low $T$. 
The two Cu2 spins in a tetramer are weakly and antiferromagnetically 
coupled to each other 
through a Cu1-Cu1 dimer in the same tetramer. 
Thus, the magnetization of Cu2 spins is large. 
The susceptibility and magnetization of CuInVO$_5$ 
resemble those of Cu$_3$(P$_2$O$_6$OH)$_2$,  
which has spin-$\frac{1}{2}$ trimerized chains 
expressed as the sequence -Cu(1)-Cu(2)-Cu(2)- \cite{Hase06,Hase07}.
The AF exchange interaction is largest 
between two neighboring Cu(2) spins (111 K). 
The magnetization of Cu(2) spins is small at low $T$. 
In each chain, two Cu(1) spins are weakly coupled to each other 
through an intermediate Cu(2)-Cu(2) AF dimer. 
The magnetization of Cu(1) spins is large.  

In CuInVO$_5$, the low-energy triplet excitation is expected 
to have a finite gap above $T_{\rm N}$ 
as in Cu$_2$CdB$_2$O$_6$ \cite{Hase15}. 
When the temperature is decreased, 
the gap closes at $T_{\rm N}$ and 
the triplet excitation splits into 
a longitudinal mode and twofold degenerate transverse modes 
at $T<T_{\rm N}$.
Slightly below $T_{\rm N}$, 
the ordered moment is small and 
the longitudinal mode is expected 
to be in the low-energy region (on the order of 1 meV).
Thus, the ordered phase in CuInVO$_5$ 
corresponds to the pressure-induced ordered phase 
in TlCuCl$_3$ \cite{Tanaka03,Oosawa04,Ruegg08,Merchant14,Matsumoto04}
and KCuCl$_3$ \cite{Goto06,Kuroe12}. 
CuInVO$_5$ may be useful for studying 
the longitudinal mode under the atmospheric pressure.

The magnetic structure is necessary 
to calculate magnetic excitations. 
In future, we will determine the magnetic structure of CuInVO$_5$
by neutron powder diffraction experiments. 
Although $^{115}$In atoms (natural abundance 95.7 \%) 
strongly absorb neutrons 
[the thermal absorption cross section is 202(2) barn for 0.0253 eV], 
it is expected to be possible to obtain diffraction patterns 
to determine the magnetic structure 
using a double-wall container. 
It is difficult to observe magnetic excitations
by INS experiments 
because of the strong neutron absorption by $^{115}$In atoms. 
We intend to form single crystals of CuInVO$_5$ and 
perform Raman scattering experiments on them. 
We expect to observe one-magnon Raman scattering 
indicating longitudinal-mode magnetic excitations 
as in TlCuCl$_3$ \cite{Kuroe08} 
and KCuCl$_3$ \cite{Kuroe12}. 

\section{Conclusion}

We measured the temperature dependence of 
the magnetic susceptibility and specific heat 
and the magnetic-field dependence of 
the magnetization of CuInVO$_5$. 
An antiferromagnetically ordered state 
appears below $T_{\rm N} = 2.7$ K. 
We observed a $\frac{1}{2}$ quantum magnetization plateau 
above 30 T at 1.3 K. 
An isolated antiferromagnetic spin-$\frac{1}{2}$ tetramer model 
with $J_1 = 240$ and $J_2 = -142$ K 
can closely reproduce the magnetic susceptibility above 30 K. 
We were able to explain the magnetization curves 
using the interacting spin tetramer model with
the effective intertetramer interaction $J_{\rm eff} = 30$ \ K.
The value of the spin gap (singlet-triplet gap) is 17 K (1.5 meV)
in the isolated spin tetramer. 
Detectable low-energy (on the order of 1 meV) longitudinal-mode 
magnetic excitations 
may exist in CuInVO$_5$. 

\begin{acknowledgments}

This work was financially supported by 
Japan Society for the Promotion of Science (JSPS) 
KAKENHI (Grant Nos. 23540396 and 15K05150) and 
by grants from National Institute for Materials Science (NIMS). 
M. Matsumoto was supported by 
JSPS KAKENHI (Grant No. 26400332). 
The high-field magnetization experiments were conducted 
under the Visiting Researcher's Program of 
the Institute for Solid State Physics (ISSP), the University of Tokyo. 
We are grateful to S. Matsumoto 
for sample syntheses and x-ray diffraction measurements.   

\end{acknowledgments}

\newpage 

\begin{references}

\bibitem{Tanaka03}
H. Tanaka, K. Goto, M. Fujisawa, T. Ono, and Y. Uwatoko, 
Magnetic ordering under high pressure 
in the quantum spin system TlCuCl$_3$, 
Physica B {\bf 329-333}, 697 (2003).

\bibitem{Oosawa04}
A. Oosawa, K. Kakurai, T. Osakabe, M. Nakamura, 
M. Takeda, and H. Tanaka,
Pressure-induced successive magnetic phase transitions in the
spin gap system TlCuCl$_3$, 
J. Phys. Soc. Jpn. {\bf 73}, 1446 (2004).

\bibitem{Oosawa99}
A. Oosawa, M. Ishii, and H. Tanaka,  
Field-induced three-dimensional magnetic ordering 
in the spin-gap system TlCuCl$_3$,  
J. Phys.: Condens. Matter. {\bf 11}, 265 (1999).

\bibitem{Nikuni00}
T. Nikuni, M. Oshikawa, A. Oosawa, and H. Tanaka, 
Bose-Einstein condensation of dilute magnons in TlCuCl$_3$, 
Phys. Rev. Lett. {\bf 84}, 5868 (2000).

\bibitem{Tanaka01}
H. Tanaka, A. Oosawa, T. Kato, H. Uekusa, Y. Ohashi, 
K. Kakurai, and  A. Hoser, 
Observation of Field-Induced Transverse N\'eel Ordering 
in the Spin Gap System TlCuCl$_3$, 
J. Phys. Soc. Jpn. {\bf 70}, 939 (2001).

\bibitem{Goto06}
K. Goto, M. Fujisawa, H. Tanaka, Y. Uwatoko, A. Oosawa, 
T. Osakabe, and K. Kakurai, 
Pressure-Induced Magnetic Quantum Phase Transition 
in Gapped Spin System KCuCl$_3$,  
J. Phys. Soc. Jpn. {\bf 75}, 064703 (2006).

\bibitem{Oosawa02}
A. Oosawa, T. Takamasu, K. Tatani, H. Abe, N. Tsujii, O. Suzuki, 
H. Tanaka, G. Kido, and K. Kindo, 
Field-induced magnetic ordering in the quantum spin system KCuCl$_3$,   
Phys. Rev. B {\bf 66}, 104405 (2002).

\bibitem{Kuroe08}
H. Kuroe, K. Kusakabe, A. Oosawa, T. Sekine, F. Yamada, H. Tanaka, and M. Matsumoto,
Magnetic field-induced one-magnon Raman scattering 
in the magnon Bose-Einstein condensation phase of TlCuCl$_3$,
Phys. Rev. B {\bf 77}, 134420 (2008).

\bibitem{Ruegg08}
Ch. R\"uegg, B. Normand, M. Matsumoto, A. Furrer, D. F. McMorrow, 
K. W. Kr\"amer, H.-U. G\"udel, S. N. Gvasaliya, H. Mutka, and M. Boehm, 
Quantum Magnets under Pressure: Controlling Elementary Excitations 
in TlCuCl$_3$, 
Phys. Rev. Lett. {\bf 100}, 205701 (2008).

\bibitem{Merchant14}
P. Merchant. B. Normand, K. W. Kr\"amer, M. Boehm, 
D. F. McMorrow, and Ch. R\"uegg, 
Quantum and classical criticality in a dimerized
quantum antiferromagnet, 
Nat. Phys. {\bf 10}, 373 (2014).

\bibitem{Kuroe12}
H. Kuroe, N. Takami, N. Niwa, T. Sekine, M. Matsumoto, 
F. Yamada, H. Tanaka, and K. Takemura,
Longitudinal magnetic excitation in KCuCl$_3$ studied 
by Raman scattering under hydrostatic pressures,
J. Phys.: Conf. Ser. {\bf 400}, 032042 (2012).

\bibitem{Matsumoto04}
M. Matsumoto, B. Normand, T. M. Rice, and M. Sigrist, 
Field- and pressure-induced magnetic quantum phase transitions 
in TlCuCl$_3$, 
Phys. Rev. B {\bf 69}, 054423 (2004).

\bibitem{Matsumoto08a}
M. Matsumoto, H. Kuroe, A. Oosawa, and T. Sekine, 
One-Magnon Raman Scattering as a Probe of Longitudinal Excitation Mode
in Spin Dimer Systems, 
J. Phys. Soc. Jpn. {\bf 77}, 033702 (2008).

\bibitem{Goldstone62}
J. Goldstone, A. Salam, and S. Weinberg, 
Broken Symmetries, 
Phys. Rev. {\bf 127}, 965 (1962).

\bibitem{Higgs64}
P. W. Higgs, 
Broken Symmetries and the Masses of Gauge Bosons, 
Phys. Rev. Lett. {\bf 13}, 508 (1964).

\bibitem{Sachdev11}
S. Sachdev, 
Quantum Phase Transitions Second Edition 
(Cambridge University Press, Cambridge, U.K., 2011).

\bibitem{Podolsky11}
D. Podolsky, A. Auerbach, and D. P. Arovas, 
Visibility of the amplitude (Higgs) mode in condensed matter, 
Phys. Rev. B {\bf 84}, 174522 (2011).

\bibitem{Matsumoto10}
M. Matsumoto, H. Kuroe, T. Sekine, and T. Masuda, 
Transverse and Longitudinal Excitation Modes in Interacting Multispin Systems, 
J. Phys. Soc. Jpn. {\bf 79}, 084703 (2010).

\bibitem{Hase05}
M. Hase, M. Kohno, H. Kitazawa, O. Suzuki, K. Ozawa, 
G. Kido, M. Imai, and X. Hu, 
Coexistence of a nearly spin-singlet state and antiferromagnetic long-range order 
in quantum spin system Cu$_2$CdB$_2$O$_6$, 
Phys. Rev. B {\bf 72}, 172412 (2005).

\bibitem{Hase09}
M. Hase, A. D\"onni, V. Yu. Pomjakushin, L. Keller, 
F. Gozzo, A. Cervellino, and M. Kohno, 
Magnetic structure of Cu$_2$CdB$_2$O$_6$ exhibiting 
a quantum-mechanical magnetization plateau and 
classical antiferromagnetic long-range order, 
Phys. Rev. B {\bf 80}, 104405 (2009).

\bibitem{Hase15}
M. Hase, K. Nakajima, S. Ohira-Kawamura, Y. Kawakita, T. Kikuchi, and 
M. Matsumoto, 
Magnetic excitations in the spin-$\frac{1}{2}$ tetramer substance 
Cu$_2$$^{114}$Cd$^{11}$B$_2$O$_6$ 
obtained by inelastic neutron scattering experiments, 
Phys. Rev B {\bf 92}, 184412 (2015).

\bibitem{Masuda04}
T. Masuda, A. Zheludev, B. Grenier, S. Imai, K. Uchinokura, 
E. Ressouche, and S. Park, 
Cooperative Ordering of Gapped and Gapless Spin Networks 
in Cu$_2$Fe$_2$Ge$_4$O$_{13}$, 
Phys. Rev. Lett. {\bf 93}, 077202 (2004).

\bibitem{Masuda05}
T. Masuda, A. Zheludev, B. Sales, S. Imai, K. Uchinokura, and S. Park, 
Magnetic excitations in the weakly coupled spin dimers and chains material 
Cu$_2$Fe$_2$Ge$_4$O$_{13}$, 
Phys. Rev. B {\bf 72}, 094434 (2005).

\bibitem{Masuda07}
T. Masuda, K. Kakurai, M. Matsuda, K. Kaneko, and N. Metoki, 
Indirect magnetic interaction mediated by a spin dimer 
in Cu$_2$Fe$_2$Ge$_4$O$_{13}$, 
Phys. Rev. B {\bf 75}, 220401(R) (2007).

\bibitem{Masuda09}
T. Masuda, K. Kakurai, and A. Zheludev, 
Spin dimers in the quantum ferrimagnet Cu$_2$Fe$_2$Ge$_4$O$_{13}$ 
under staggered and random magnetic fields, 
Phys. Rev. B {\bf 80}, 180412(R) (2009).

\bibitem{Hase97}
M. Hase, K. M. S. Etheredge, S.-J. Hwu, K. Hirota, and G. Shirane, 
Spin-singlet ground state with energy gaps in Cu$_2$PO$_4$: 
Neutron-scattering, magnetic-susceptibility, and ESR  measurements, 
Phys. Rev. B {\bf 56}, 3231 (1997). 
In this reference, the Hamiltonian is defined as 
${\cal H} = \sum_{i,j} 2J_{ij} S_i \cdot S_j$ instead of 
${\cal H} = \sum_{i,j} J_{ij} S_i \cdot S_j$ in the present paper.

\bibitem{Moser99}
P. Moser, V. Cirpus, and W. Jung, 
CuInOVO4 - Single Crystals of a Copper(II) Indium Oxide Vanadate 
by Oxidation of Cu/In/V Alloys, 
Z. Anorg. Allg. Chem. {\bf 625}, 714 (1999). 

\bibitem{Comment1}
The estimation of the magnetic specific heat strongly depends on 
the estimation of the lattice specific heat. 
There is no nonmagnetic isostructural compound for CuInVO$_5$. 
The magnetic specific heat probably remains up to a high $T$ 
because of the low-dimensional spin system. 
We cannot determine whether 
the magnetic specific heat estimated 
on the basis of an assumption of the lattice specific heat 
is correct or not. 
Accordingly, we did not estimate the magnetic specific heat.  

\bibitem{Hase93a}
M. Hase, I. Terasaki, and K. Uchinokura, 
Observation of the Spin-Peierls Transition 
in Linear Cu$^{2+}$ (Spin- $\frac{1}{2}$) 
Chains in an Inorganic Compound CuGeO$_3$,
Phys. Rev. Lett. {\bf 70}, 3651 (1993).

\bibitem{Hase93b}
M. Hase, I. Terasaki, Y. Sasago, K. Uchinokura, and H. Obara, 
Effects of Substitution of Zn for Cu in the Spin-Peierls Cuprate, CuGeO$_3$: 
The Suppression of the Spin-Peierls Transition and 
the Occurrence of a New Spin-Glass State,
Phys. Rev. Lett. {\bf 71}, 4059 (1993).

\bibitem{Hase93c}
M. Hase, I. Terasaki, K. Uchinokura, M. Tokunaga, 
N. Miura, and H. Obara, 
Magnetic phase diagram of the spin-Peierls cuprate CuGeO$_3$, 
Phys. Rev. B {\bf 48}, 9616 (1993).

\bibitem{Hase06}
M. Hase, M. Kohno, H. Kitazawa, N. Tsujii, O. Suzuki, K. Ozawa, 
G. Kido, M. Imai, and X. Hu, 
1/3 magnetization plateau observed in the spin-1/2 trimer chain compound 
Cu$_3$(P$_2$O$_6$OH)$_2$, 
Phys. Rev. B {\bf 73}, 104419 (2006).

\bibitem{Hase07}
M. Hase, M. Matsuda, K. Kakurai, K. Ozawa, H. Kitazawa, N. Tsujii, 
A. D\"onni, M. Kohno, and X. Hu, 
Direct observation of the energy gap generating 
the 1/3 magnetization plateau 
in the spin-1/2 trimer chain compound Cu$_3$(P$_2$O$_6$OD)$_2$ 
by inelastic neutron scattering measurements, 
Phys. Rev. B {\bf 76}, 064431 (2007).

\end{references}

\end{document}